%% file: inv_ps.tex
\documentclass[12pt]{article}
\usepackage{amsmath,amsfonts,amssymb,a4,color,graphics}
\usepackage[latin1]{inputenc}

\newcommand{\R}{{\mathbb{R}}}

\newcommand{\Z}{{\mathbb{Z}}}
\newcommand{\N}{{\mathbb{N}}}

\def\ha{\frac{1}{2}}

\def\pa{\partial}
\def\ra{\rightarrow}
\def\ga{\alpha}

\def\gd{\delta}
\def\ge{\varepsilon}

\def\gf{\varphi}
\def\gg{\gamma}

\def\gl{\lambda}
\def\go{\omega}

\def\gr{\rho}

\newtheorem{defi}{Definition}[section]

\newtheorem{lemm}{Lemma}[section]

\newtheorem{rem}{Remark}[section]

\newtheorem{theo}{Theorem}[section]

\newenvironment{demo}{\noindent {\it Proof.--}
      \begin{quotation}\noindent}{\end{quotation}\hfill$\square $}

\begin{document}

\title{A semi-classical inverse problem II:\\
reconstruction of the potential}
\author{Yves  Colin de Verdi\`ere\footnote{Grenoble University, 
Institut Fourier,
 Unit{\'e} mixte
 de recherche CNRS-UJF 5582,
 BP 74, 38402-Saint Martin d'H\`eres Cedex (France);
{\color{blue} {\tt yves.colin-de-verdiere@ujf-grenoble.fr}}}}


\maketitle

\section{Introduction}

This paper is  the continuation of \cite{YV}, where Victor Guillemin
and
I  proved the
following result:
the Taylor expansion of the potential $V(x)$ $(x\in\R) $ at a non
degenerate
critical point  $x_0$ of $V$, satisfying $V'''(x_0)\ne 0$,
 is determined by the semi-classical spectrum of the
associated
Schr\"odinger operator near the corresponding critical value $V(x_0)$.
Here, I prove  results which are  stronger in some aspects: the 
potential itself, without any analyticity assumption, but with
 some genericity conditions, 
 is determined
 from the semi-classical spectrum. Moreover, our method gives an
 explicit way to reconstruct the potential.

Inverse spectral results for Sturm-Liouville operators
are due to Borg, Gelfand, Levitan, Marchenko and others
(see for example \cite{L-G}).
They need  the  spectra of the differential operator
with two different boundary conditions 
in order to recover the potential. Our results are
different in several   aspects:
\begin{itemize}
\item They are local using only the part of the spectrum
included in some interval $]-\infty , E[ $  in order to get $V$ in
the inverse image $\{ x |V(x)< E \}$  of this interval.
\item They need only approximate spectra. 
\item They still apply if the operator is essentially self-adjoint. 
\end{itemize}

After having completed the present  work, I founded
that similar methods were already used by David Gurarie \cite{Gur}
in order to recover a surface of revolution from the joint
spectrum of the Laplace operator and the momentum operator $L_z$.
Our genericity assumptions are weaker and more explicit:
\begin{itemize}
\item 
David Gurarie assumes that the potential is a Morse function
with pairwise different critical values,
while we assume only a weak non degeneracy condition
 (see Section \ref{ss:crit}).
\item His  argument for the separation of spectra associated to
the different wells is less explicit than ours which
uses the 
semi-classical trace formula (see Section \ref{sec:sep}).
\item He does not say a word about the problem of a non generic
symmetry defect and explicit  non isomorphic 
potentials with the same semi-classical spectra
(Section \ref{sec:contrex} and Assumption 3 in Theorem \ref{theo:main}).
\end{itemize}

For a recent review  on the use of semi-classics in
  inverse spectral problems,
 the 
reader could look at \cite{Ze}. 

\section{Motivation I:  surfaces of revolution}

Let us consider a surface of revolution with a metric
\[ ds^2=dx^2 + a^4 (x) dy^2 \]
with $x \in [0,L]$ and $ y\in \R / 2\pi \Z $.
We  assume that $a(0)=a(L)=0$,  $a(x)>0$ for $0<x<L$ and 
$a$ is smooth.
The volume element is given by 
$dv= a^2 (x) |dx dy| $.
The Laplace operator is:
\[ \Delta =-\frac{\pa ^2 }{\pa x^2} 
-\frac{2a'}{a}\frac{\pa  }{\pa x}
-\frac{1}{a^4}\frac{\pa ^2 }{\pa y^2}  \ .  \] 
Using the change of function
$f=Fa$,  we get
the operator
$P=a\Delta a ^{-1}$ which is formally symmetric w.r. to 
$|dx dy |$:
\[P= -\frac{\pa ^2 }{\pa x^2} +\frac{a''}{a}
   -\frac{1}{a^4}\frac{\pa ^2 }{\pa y^2}\ .\]
If $F(x, y)=\gf (x ){\rm exp}(il y)$ with $l\in  \Z$,
we define $Q_l$ as follows 
\[ PF=l^2 (Q_l \gf )e^{ily} \ ,\] 
and puting  $\hbar =l^{-1}$, we get
\[ Q_\hbar  \gf =-\hbar ^2 \gf'' +\left( a^{-4}+ \hbar ^2 W \right) \gf \ \]
with 
$ W=\frac{a''}{a} $.
It implies that the knowledge of the joint spectrum of $\Delta $
and $\pa _y$ is closely related to the spectra of
$Q_\hbar $ for $\hbar =1/l $ with $l\in \Z \setminus 0$.
This relates our paper to Gurarie's result \cite{Gur}.

\section{Motivation II:  effective surface waves Hamiltonian}
\label{sec:acoustic}

In our paper \cite{CV1}, we started  with the following acoustic 
wave equation\footnote{$u=u({\bf x},z,t)$ is the
 pressure, $n=K/\gr $ with $\gr $
the density and $K>0$ the incompressibility assumed to be a  constant. The
acoustic
wave equation is a simplification of the elastic wave equation which
holds
if the medium is fluid.} 
\begin{equation}\label{equ:ondes}
\left\{ \begin{array}{l}
u_{tt}-{\rm div}(n~{\rm grad}u)=0\\
u({\bf x},0,t)=0 
\end{array} \right.
\end{equation}
in the half space $X=\R^{d-1}_{\bf x}\times ]-\infty , 0 ]_z$
where 
 $n(z): \R _- \ra \R _+$ is 
 a non negative  function which 
satisfies
\[ 0< n_0:=\inf n(z)< n_\infty:= \liminf _{z\ra -\infty }n(z)~.\]
This equation  describes the propagation of acoustic waves in a medium
which is stratified: the variations of the density  are on much
smaller scales vertically than horizontally\footnote{In \cite{CV1},
we took  a more complicated function $n({\bf x},z)=N({\bf x},z/\ge, z) $
with $N$ smooth and $\ge $ small}.
This equation admits solutions of the
form ${\rm exp}(i(\go t -{\bf x}{\bf \xi}))v(z) $
provided that $v$ is an eigenfunction 
of the operator  
$L_{{\bf \xi} }$ on the half line $z\leq 0$ defined as follows:
\begin{equation} \label{equ:sturm}
L_{{{\bf \xi} }}v:=-\frac{d}{dz}
\left(n(z)\frac{dv}{dz}\right) + n(z)| {\bf \xi} | ^2 v 
\end{equation}
with Dirichlet boundary conditions and eigenvalue $\go^2$.
These solutions are exponentially localized near the boundary
provided that $\go^2 $ is in the discrete spectrum of $L_{{\bf \xi} }$
contained in $J:=]n_0 |{\bf \xi} |^2, n_\infty |{\bf \xi} |^2 [$.

Let us denote by
 $\gl_1 ({\bf \xi})< \gl_2({\bf \xi})< \cdots <\gl_j({\bf \xi})<
\cdots $ the spectrum of $L_{{{\bf \xi} }}$ in the interval $J$
 and $v_j({\bf \xi},z)$
the associated normalized eigenfunctions.
The unitary map from $L^2(\pa X)$ into $L^2(X)$ defined by
\[ T_j (a):=(2\pi)^{-(d-1)}\int
_{\R^{d-1}}\hat{a}({\bf \xi})v_j({\bf \xi},z)e^{i{\bf x}{\bf \xi} }d{\bf \xi} ~,\]
with $\hat{a}({\bf \xi}):=\int _{\R^{d-1}}a({\bf x})e^{-i{\bf x}{\bf \xi}} d{\bf x} $,
satisfies:
\[ P T_j=T_j {\rm Op}(\gl_j ) ~,\]
where $P=-{\rm div}(n~{\rm grad}u) $ with Dirichlet boundary conditions
and  ${\rm Op}(\gl_j )$  is an elliptic  pseudo-differential operator 
of degree $2$ and  
of symbol $\gl_j$.
So that, for each $j=1,\cdots $, we get an effective
surface wave Hamiltonian with the Hamiltonian $\gl_j$.
The map $T:\oplus  _{j=1}^\infty  L^2(\pa X) \ra L^2(X)$ given
by $T=\oplus _{j=1}^\infty T_j $ is an injective isometry.

We see that the high frequency surface waves are 
associated to the semi-classical spectrum of a Schr\"odinger type 
operator
\[{\cal L}_{\hbar } = -\hbar ^2 \frac{d}{dz}
\left(n(z)\frac{d}{dz}\right) + n(z) ~,\]
with $\hbar = \| {\bf \xi} \| ^{-1} $.
One can try to recover $n(z)$ from
the propagation of surface waves: this is equivalent to 
get the  operator ${\cal L}_{\hbar} $  from its semi-classical spectrum. 

\section{Some notations}\label{sec:not}
The following notations will be used everywhere in this paper.
The {\it interval} $I$ is defined by
$I=]a,b[$ with $-\infty \leq a<b \leq +\infty$.
The {\it potential}  $V:I=]a,b[\ra \R $ is a smooth function with
$-\infty <E_0:=\inf V < E_\infty =\liminf_{x\ra \pa I} V(x) $.
We will denote by 
$\hat{H}$
any self-adjoint extension of the operator $-\hbar^2 \frac{d^2}{dx^2}+V(x) $
defined on $C_o^\infty (I)$.
The discrete spectrum of $\hat{H}_\hbar $ will be denoted
by 
\[(E_0<) \gl_1 (\hbar)<  \gl_2 (\hbar)<\cdots < \gl_l (\hbar)<\cdots ~.\]
The semi-classical limit is associated to the classical
Hamiltonian $H=\xi^2 + V(x)$ and the dynamics
$dx/dt=\xi,~d\xi/dt =-V'(x)$.

\begin{defi} \label{def:dist}
We say that $\mu_l(\hbar ) $ is a  semi-classical spectrum
of $\hat{H}$ mod $o(\hbar ^N)$
in  $[E_0, E]$ if, for any $F< E$,
\[\left(  \sum _{\gl_l (\hbar) \leq F}|\gl_l(\hbar)
 -\mu_l(\hbar) |^2 \right)^\ha =
o(\hbar ^{N-\ha}) ~.\]
\end{defi}
If we have a uniform approximation of the eigenvalues
up to $o(\hbar ^{N})$, it is also a
  semi-classical spectrum
of $\hat{H}$ mod $o(\hbar ^N)$ in the previous $l^2$ sense
because the number of eigenvalues in $]-\infty ,F]$
is $O(\hbar^{-1})$.

\section{A Theorem for  one well potentials}
\label{sec:1well}

\begin{theo} \label{theo:main}
Let us assume that the potential 
    $V:I\ra \R$
 satisfies:
\begin{enumerate}
\item {\bf A single well below $E$:}
there exists $E \leq E_\infty $ so that, for any  $y\leq  E $,  
the sets $I_y:=\{ x|V(x)\leq y \}$ 
are  connected. The intervals $I_y$ are  compact for $y<E$.
There exists a unique $x_0$ so that
$V(x_0)=E_0~(=\inf_{x\in I} V(x))$.
 For any $y$ with $E_0< y \leq E$, if the interval $I_y$
is defined by   $I_y=[f_-(y),f_+(y)]$, we have 
$V'(x_0)=0$, 
$V'(x)<0$ for $f_-(E)< x <x_0 $
and
$V'(x)>0$ for $x_0<x<f_+(E) $.
\item  {\bf A genericity hypothesis at the minimum:}
there exists $N\geq 2$ so that 
 the $N$-th derivative $V^{(N)}(x_0)$ does not vanish.
\item {\bf A generic  symmetry defect:}
 if there exists $x_\pm $, satisfying
$f_-(E)<x_-< x_+
< f_+(E)$ and  $\forall n\in \N,~V^{(n)}(x_-)=(-1)^n V^{(n)} (x_+)$,
  then $V$ is globally
even w.r. to $x_0=(x_-+x_+)/2$ on the interval
$I_E$.
 This is true for example if $V$ is real analytic.
\end{enumerate}
Then the {\rm spectra  modulo $o(\hbar^2)$} in the interval
$]-\infty , E[ $ of the Schr\"odinger operators
$\hat{H}_\hbar $,
for a sequence $\hbar _j \ra 0^+$,
 determine  $V$ in the interval
$I_E $  up to a symmetry-translation  $V(x) \ra
V(c\pm x) $.
\end{theo}
\begin{figure}[hbtp]
  \begin{center}
    \leavevmode
    \input{potential.pstex_t}
    \caption{The potential $V$ and the functions $f_+$ and
$f_-$}
    \label{fig:pot1}
  \end{center}
\end{figure}
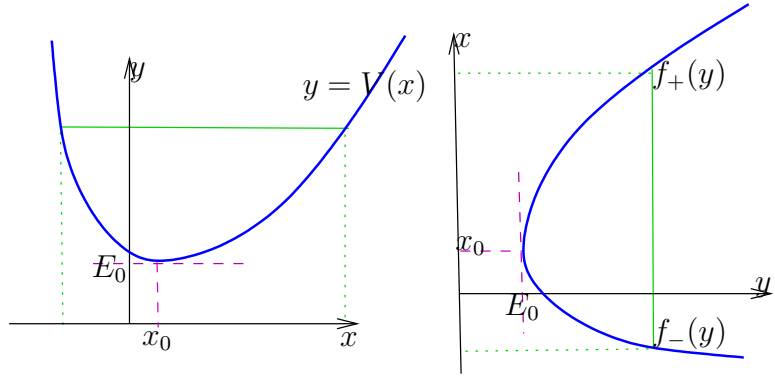

\section{One well potentials : Bohr-Sommerfeld rules
and a $\Psi DO$ trace formula}\label{sec:bs}
From \cite{CV2}, we know that the semi-classical spectrum
(i.e. the spectrum up to $O(\hbar^\infty)$)
of $\hat{H}_\hbar$ in the interval 
$]E_0, E[$
is given by 
\[ \Sigma (\hbar)=\{y~|~E_0 < y<E {\rm ~and~}
S(y) \in 2\pi \hbar \Z \} \]
 where, for $E_0<y<E$, the function $S$ admits the
formal series expansion
$S(y)\equiv S_0(y)+ \hbar \pi + \hbar ^2 S_2(y) +\hbar ^4 S_4 (y) +\cdots  $
(the formal series $S$ will be called the {\it semi-classical action}
and the remainder term in the expansion is uniform in every
compact sub-interval of $]E_0,E[$)
with 
\begin{itemize}
\item
$S_0 (y)=\int _{\gg _y}\xi dx $
with $\gg _y=\{ (x,\xi)|H(x,\xi)=y \}$
oriented according to the classical dynamics 
and 
\[ \frac{d S_0}{dy}(y)=\int _{f_-(y)}^{f_+(y)} \frac{dx}{\sqrt{y-V(x)}}~\]
is the {\it period~} $T(y)$ of the trajectory of energy $y$ for the
classical Hamiltonian
$H$,
\item If $t$ is the time parametrization of $\gg_y$,
\[ S_2(y)=-\frac{1}{12}\frac{d}{dy}\int _{\gg_y} V''(x) dt~,\]
which can be rewritten as:
\[ S_2(y)=-\frac{1}{12}\frac{d}{dy}\left( \int _{f_-(y)}^{f_+(y)}
\frac{V''(x)dx }{\sqrt{y-V(x)}}~\right).\]
\item For $j\geq 1$, $S_{2j}(y)$ is a linear combination
 of  expressions of the form
\[ \left( \frac{d}{dy} \right) ^n \int _{\gg_y}P(V',V'',\cdots )
dt~,\] where
 $dt$ is the differential of the time on $\gg_y$:
outside the caustic set $dt=dx/2\xi$.
\end{itemize}

In what follows, we will use only $S_0$ and $S_2$.
It will be convenient to relate the semi-classical action 
to the spectra by using the following trace formula:
\begin{theo}\label{theo:opdtrace} {\bf ($\Psi DO$ trace formula)}
Let $f\in C_o^\infty (]E_0,E[)$
 and $F(y):=-\int _y^\infty f(u)du $,
we have, with $Z=T^\star I$: 
\[ {\rm Trace}F(\hat{H})=\frac{1}{2\pi \hbar }
\left( \int _{Z}F(H) dx d\xi 
+  {\hbar ^2}\int _{E_0}^E f(y) (S_2 (y)+\hbar^2 S_4(y)+ \cdots) dy  \right) + O(\hbar ^\infty)~.\]
This formula implies that $S_0$ and $S_2$ are determined 
by the semi-classical spectrum mod $o(\hbar ^2)$ in $]-\infty, E[$.
\end{theo}
This Theorem is closely related to
(but a bit stronger) than what is
proved in  my paper \cite{CV2}. The trace formula 
 contains implicitely the 
Maslov index.

\section{Two potentials with the same semi-classical spectra}
\label{sec:contrex}
\begin{figure}[hbtp]
  \begin{center}
    \leavevmode
\input{contrex.pstex_t}
    \caption{The (graphs of the) two
 potentials are the same in the sets $II$ and
$III$, they are mirror image of each other in $I$
(green curve and dotted green curve), the potential
is even  in the set  $II$.}
    \label{fig:contrex}
  \end{center}
\end{figure}
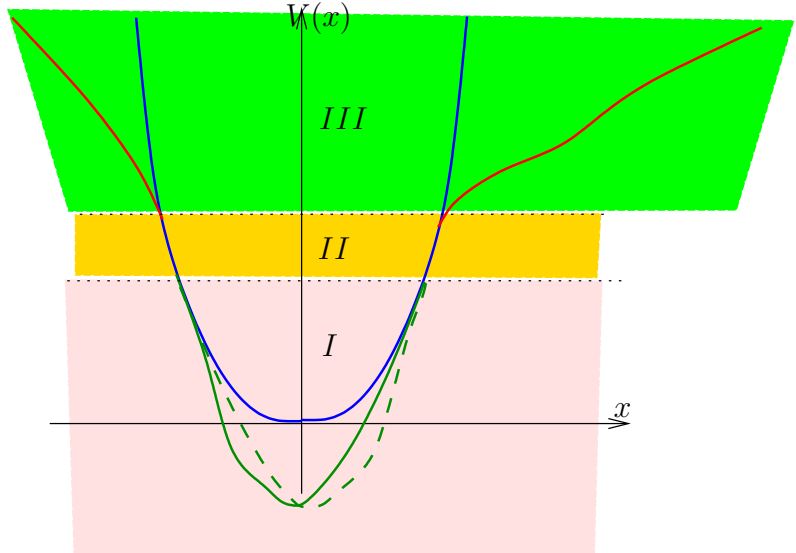

We introduced a genericity Assumption 3 on symmetry defects
 in Theorem \ref{theo:main}.
The Figure  \ref{fig:contrex} ~ shows two  one well potentials with the same
semi-classical spectra mod $O(\hbar ^\infty)$.
The fact that they have the same semi-classical spectra comes from
the description of Bohr-Sommerfeld rules in  Section \ref{sec:bs}.
 
 It would be nice to prove that they do NOT
have the same spectra!

\section{One well potentials : the proof of Theorem \ref{theo:main}}

\subsection{Some useful Lemmas}
\begin{lemm}
The semi-classical spectra modulo $o(\hbar ^2)$ in $]E_0,E[$ determine
the  actions $S_0 (y)$ and $S_2(y)$ for $y\in ]E_0,E[ $.
\end{lemm}
It is a consequence of Theorem \ref{theo:opdtrace}.

\begin{lemm}\label{lemm:limite}
If $V$ satisfies Assumption 2 in Theorem \ref{theo:main}, we have:
\[ \lim _{y\ra E_0} \int _{\gg_y} V''(x) dt 
=\pi \sqrt{2V''(x_0)} ~.\]
This holds  even if the minimum is degenerate\footnote{I do not know
if this is still true without the genericity Assumption 2
in Theorem \ref{theo:main}; it is the
only place where I use it}.
\end{lemm}

The Lemma is clear if $V''(x_0)>0$: the limit is then $V''(x_0)$ times
 the period of small oscillations of a pendulum which is
$\pi / \sqrt{2/V''(x_0)}$.

 Let us consider the case of an
isolated degenerate minimum with $V(x)=E_0+ a(x-x_0)^N (1+o(1))$ ($a>0,~N>2$),
we can check that the integral to be evaluated is
$O\left((y-E_0)^{\frac{3}{2}-\frac{3}{N}}\right)=o(1)$.

\begin{lemm} \label{lemm:vp}
We have
\[ \lim _{y\ra 0}\left( \frac{1}{f'_+(y)} -\frac{1}{f'_-(y)} \right)=
0 ~.\]
\end{lemm}

\begin{lemm}\label{lemm:first}
If $x_0$ is the unique point where 
$V(x_0)=\inf V =E_0$,
the first eigenvalue of $\hat{H}_\hbar $ satisfies
$\gl _1 (\hbar )=E_0 + \hbar \sqrt{V''(x_0)/2}  + o(\hbar ) $
\end{lemm}
This is well known if $V''(x_0)>0 $ and is still true
otherwise by comparison: if $E_0 \leq V(x) \leq A(x-x_0)^2 $ with
$A>0$, 
near $x_0$ then $E_0< \gl_1(\hbar )\leq 2\pi  \hbar \sqrt{A}$.
\subsection{Rewriting $V$ using $F$ and $G$}
We will denote by $F=\ha (f_++f_-) $
and $G=\ha (f_+-f_- )$.
\begin{itemize}
\item 
The function $F$ is smooth on $]E_0,E[$, continuous on $[E_0, E[$
(smooth in the non degenerate case $V''(x_0)>0$
 as a consequence of the Morse Lemma), with $F(E_0)=x_0$,
 and is constant  if and only if $V$ is even w.r. to $x_0$.
More generally, if $F$ is constant  on some interval, $V$ is even
on the inverse image of that interval.
 We call $F$  the {\it parity defect}.
\begin{lemm} \label{lemm:parity}
Under the Assumption 3 in Theorem \ref{theo:main},
the function $F'$ is determined up to $\pm $ by its
square.
\end{lemm} 
\item
The function $G$ is  smooth on $]E_0,E[$,
continuous at $y=E_0$.
We have  $G(E_0)=0$.
It is clear that, from  $F$ and $G$, we can
 recover the restriction of $V$ to
$I_E$.
\end{itemize}

\subsection{How to get $V$ from $S_0$ and
$S_2$} \label{sec:trick}

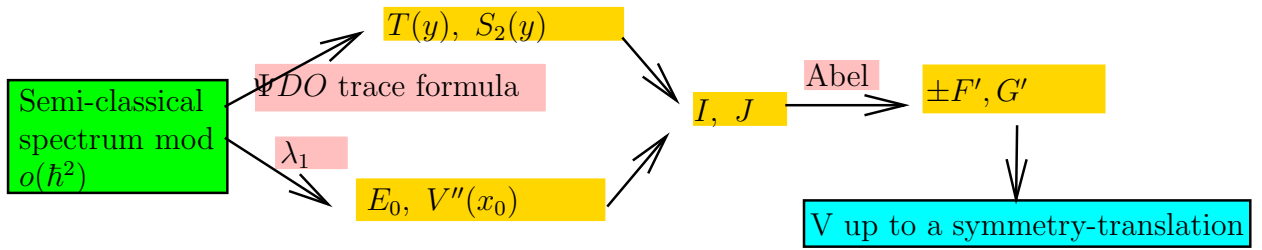
\begin{figure}[hbtp]
  \begin{center}
    \leavevmode
\input{proof.pstex_t}
    \caption{The scheme of the proof}
    \label{fig:proof}
  \end{center}
\end{figure}
Let us consider, for $E_0<y<E$,
\[ I(y):=\int _{f_-(y)}^{f_+(y)} \frac{dx}{\sqrt{y-V(x)}} \]
and 
\[ J(y)=  \int _{f_-(y)}^{f_+(y)}
\frac{V''(x)dx }{\sqrt{y-V(x)}}~.\]
We have $I(y)=dS_0(y)/dy$ and $S_2(y)=-(1/12) dJ(y)/dy$.
This implies that  $S_0$, $S_2$ and the limit $J(E_0)$
 determine $I$ and $J$.
The limit $J(E_0)$ is determined by $V''(x_0)$ (Lemma \ref{lemm:limite}) which
is determined by  the first semi-classical eigenvalue (Lemma
\ref{lemm:first}).
We can  express $I$ and $J$ using $F$ and $G$.
Using the change of variables $x=f_+(u) $ for $x>x_0$
and $x=f_-(u)$ for $x<x_0$, we get:
\[ I(y) =4\int_{E_0}^y \frac{G'(u)du}{\sqrt{y-u}} \]
\[ J(y)= \int_{E_0}^y \frac{d}{du}
\left( \frac{1}{f'_+(u)}-  \frac{1}{f'_-(u)}
\right) \frac{du}{\sqrt{y-u}} ~.\]

Using Abel's result \cite{abel} (and Appendix A), we 
can recover
$G'$ and 
\[   \frac{d}{dy} \left(  \frac{1}{f'_+(y)} -\frac{1}{f'_-(y)}
\right) =
\frac{d}{dy}\left( \frac{2G'}{G'^2 -F'^2 } \right)~. \]
Using Lemma \ref{lemm:vp}, we recover $F'^2$.
The Assumption 3 implies that there exists an unique
square root to $F'^2$ up to signs. 
From that we recover $G'$ and $\pm F'$ and hence $\pm F$ 
 and $G$ modulo constants .
This gives $V$ up to change of $x$ into $c\pm x$.

\section{Taylor expansions}

From the previous section, we see that the semi-classical
spectra determine $F'^2$ and $G$ even without assuming
the hypothesis 3 of Theorem \ref{theo:main} on symmetry defect.
It is not difficult to see that, if $V$ 
satisfies the hypothesis 2 of Theorem
\ref{theo:main},  the parity defect $F$ is a smooth function of $y^{2/N}$. 
We have the following:
\begin{lemm}
Let us give two formal powers series 
$a=\sum _{j=0}^\infty a_j t^j $
and $b=\sum _{j=0}^\infty b_j t^j $ which satisfy
$a^2=b$. The equation $f^2=b$ has exactly two solutions as formal
powers series: 
$f=\pm a$.
\end{lemm}
From this Lemma, we deduce the:
\begin{theo}
  Under the Assumptions 1 and 2 of Theorem \ref{theo:main}, but
  without Assumption 3, 
the Taylor expansion of $V$  at a  local minimum  $x_0$  is
 determined (up to mirror symmetry) by the
semi-classical spectrum modulo $o(\hbar^2)$
in a fixed neighbourhood of $E_0$.
\end{theo} 

In some aspects, this result is stronger than
the one obtained in \cite{YV}, but it requires the knowledge
of the semi-classical spectrum in a fixed neighbourhood of $E_0$,
while, in \cite{YV}, we need only $N$ semi-classical eigenvalues
in order to get $2N$ terms in the Taylor expansion.

\section{A Theorem for   a  potential with several wells}

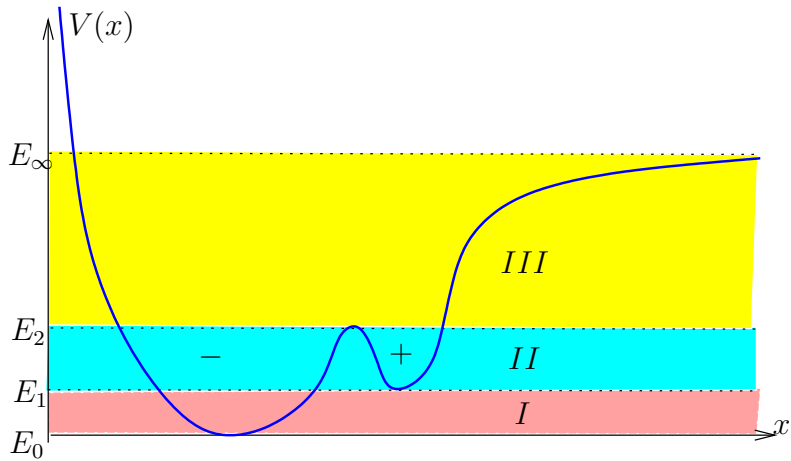
\begin{figure}[hbtp]
  \begin{center}
    \leavevmode
 \input{pot3.pstex_t}
    \caption{a 2 wells  potential $V$}
    \label{fig:pot3}
  \end{center}
\end{figure}

We will extend our main result to   cases including that 
 of Figure \ref{fig:pot3}:
a two  wells potential with three  critical values,
 $E_0=0$,  $E_1$ and $E_2$.
 We can  take any  boundary condition at $x=0$.

\subsection{The genericity Assumptions} \label{sec:technical}

{\bf In what follows, we choose $E$ so that $E_0< E \leq E_\infty$
and define  $I_E=\{ x | V(x)<E \} $.
The goal is to determine the restriction of $V$ to  $I_E$
from the semi-classical spectrum in $]-\infty, E]$.}

We need the following Assumptions which are generically satisfied.
We introduce a:
\begin{defi}
Two smooth functions $f,g:J \ra \R $
are {\rm weakly transverse} if, for every $x_0$
so that $f(x_0)=g(x_0)$, there exists an integer $N $
such that the $N$th derivative $(f-g)^{(N)}(x_0)$
does not vanish.
\end{defi}
\subsubsection{  Assumption on critical points} \label{ss:crit}
\begin{itemize}
\item for any point $x_0$ so that   $V'(x_0)=0$ and $V(x_0)<E$,
 there exists $N\geq 2$ so that,
 the $N$-th derivative $V^{(N)}(x_0)$ does not vanish.
\item 
The {\it critical values} associated
to different critical points are {\it distinct.}
\end{itemize}

{\bf The wells: }{\it 
Let us label the critical values of $V$ below $E_\infty $ 
as $E_0< E_1< \cdots < E_k < \cdots <E_\infty $
and the corresponding critical points by 
$x_0,~x_1,\cdots $. The critical values   can only
accumulate
at $E_\infty $ because the critical points are isolated and 
hence only   a  finite number of them  lies  in $\{ x |V(x)<E_\infty -c \}$
for any $c>0$.
Let us denote, for $k= 1,2,  \cdots $ by
$J_k=]E_{k-1}, E_{k}[$.
\begin{defi}
 {\rm  A well of order $k$} is a connected component
of $\{x| V(x)< E_k \} $.
\end{defi}
Let us denote by  $N_k $ the number of wells of order $k$.

For any $k$, $H^{-1}(J_k)$ is an union of $N_k$ topological
annuli $A_j^k $ and the map $H:A_j^k \ra J_k $ is a submersion whose 
fibers $H^{-1}(y)\cap A_j^k$
are topological circles $\gg_j^k (y)$  which are periodic trajectories
of the classical dynamics:
if $y\in J_k$, $H^{-1}(y)=\cup_{j=1}^{N_k} \gg_j^k (y)$.
We will denote by $T_j^k(y)=\int_{\gg_j^k}dt$,
 the corresponding classical periods.
We will often remove the index $k$ in what follows.

The semi-classical spectrum in $J_k$ is the union of $N_k$ spectra
which are given by Bohr-Sommerfeld rules associated to 
actions $S_j^k (y)$ given as in Section \ref{sec:bs}. }

\subsubsection{  A generic  symmetry defect} \label{ss:defect}
 If there exists $x_-<x_+ $, satisfying $V(x_-)=V(x_+)<E$
 and,  $\forall n\in \N,~V^{(n)}(x_-)=(-1)^n V^{(n)} (x_+)$,
  then $V$ is globally
{\it even}  on
$I_E$.

\subsubsection{ Separation of the wells}\label{ss:sep}
For any $k=1,2, \cdots $ and any $j$ with $1\leq j < l \leq N_k$,
the   classical periods $T_j(y)$ and $T_l(y) $ are weakly transverse
in $J_k$. This is assumed to hold also at $E_{k-1}$ if 
$x_k$ is a local non degenerate minimum of $V$ (in this case,
the period of the new periodic orbit is smooth at
$(E_{k-1})_+$).

\subsection{Quartic potentials}
 If $V$ is a polynomial
of degree four with two  wells like
$V(x)=x^4+a x^3 +bx^2 $ with $b<0$, the periods of the two  wells
(between
$E_1$ and 
$E_2(=0)$) are identical. This is because, on the complex projective
compactification $X_E$ (with $E<0$) 
of $\xi ^2 +V(x)=E$, the differential $dx/\xi $ is holomorphic
and the real part of $X$ consists of 2 homotopic curves in $X_E$.
One can check directly that all other actions $S_{2j}, ~j\geq 1$ coincide;
this is also proved for example in \cite{D-P} p. 191.

\subsection{The statement of the  result}

Our result is:
\begin{theo} \label{theo:main_sev}
Under the three    Assumptions in Sections \ref{ss:crit},
\ref{ss:defect} and \ref{ss:sep}, 
 $V$ is
determined in the domain  $I_E:=\{ x|V(x)< E \} $
  by the semi-classical spectrum
in $]-\infty , E[$ 
modulo $o(\hbar ^4)$ up to the following moves:
$I_E$ is an  union of open intervals $I_{E,m}$,
 each interval $I_{E,m}$  is defined up to translation and the restriction
of $V$ to each  $I_{E,m}$ is defined up to $V(x)\ra V(c-x)$ .
\end{theo}
\begin{rem} We need $o(\hbar ^4)$ in the previous Theorem while
we needed only  $o(\hbar ^2)$  in the  one well case.
This is due to the way we are able to separate
the spectra associated to the different wells.
\end{rem}

  \section{The case of several wells: the proof of Theorem
\ref{theo:main_sev}}
\subsection{What can be read from the Weyl's asymptotics?}
\begin{lemm}
Under the Assumption \ref{ss:crit},
the singular (non smooth)  points of the function 
$y\ra A(y)=\int _{H(x,\xi)\leq y}dxd\xi $
are exactly the critical values $E_0, E_1, \cdots $ of $V$.
Moreover,
\begin{itemize}
\item
 the function  $A(y)$ in smooth on $]E_k-c, E_k ]$, with $c>0$,
if and only if $x_k$ is a local minimum of $V$,
\item From the singularity of $A(y)$ at $E_k$, on can read the
value of $V''(x_k)$.
\end{itemize}
\end{lemm}

The function $A(y)$ is determined by the semi-classical
spectrum, this is a consequence of  the Weyl asymptotics:
\[ \# \{ \gl_l (\hbar)  \leq y \} \sim \frac{A(y)}{2\pi \hbar } ~.\]
This implies that the critical values $E_k $ of $V$
are determined by the semi-classical spectrum.  

\subsection{The scheme of the reconstruction}

The proof is by ``induction'' on $E$.

We start by constructing the piece of $V$ where $V(x)\leq E_1 $
using Theorem \ref{theo:main}.

We want then to construct $V$ where $E_1\leq V(x)\leq E_2$.

There are two  cases:
\begin{enumerate}
\item {\it $x_1$ is not an extremum:}
 then we are able to extend the proof
of Theorem \ref{theo:main} using the fact that we know, using Section 
\ref{sec:limites}, 
the limits of $\int _{\gg _y}V''(x) dt  $
and $f'_\pm (y) $ as $y \ra E_1^+$.
We can reduce to an Abel transform starting from $E_1$
using
\[ \int _{V(x)\leq y}=\int _{V(x)\leq E_1}+\int _{E_1\leq V(x)\leq y}\]
 where the first part is known from the knowledge of $V(x)$
in $\{x| V(x)\leq E_1 \} $.

\item {\it $x_1$ is a local minimum:} using the separation of spectra 
(Section \ref{sec:sep}) and Theorem \ref{theo:main}, we can
construct the 2 wells of order 2 if we know $V''(x_1)$. But the
 estimate 
\[ A(y)=A(E_1) + \pi \sqrt{2/V''(x_1)}(y-E_1)_+ +a(y-E_1)+ o(y-E_1)~\]
shows that  the singularity of $A(y)$ at $y=E_1$
 determines
$V''(x_1)$.

\end{enumerate}

 We then proceed to the interval $[E_2, E_3]$. A new case arises
when  {\it $x_2$ is a local maximum.} Then we need to glue together
the wells of order $2$. This case works then as before.

\subsection{Separation of spectra} \label{sec:sep}

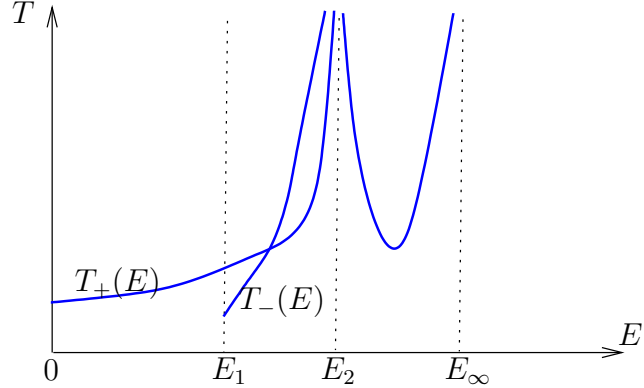
\begin{figure}[hbtp]
  \begin{center}
    \leavevmode
\input{period.pstex_t}
    \caption{The primitive periods as functions of $y$
for the Example of Figure  \ref{fig:pot3} }
    \label{fig:period}
  \end{center}
\end{figure}

Let us start with a:
\begin{lemm}\label{lemm:vanish}
Let us give some open interval $J$
and assume that we have 
a function 
\[ F(x)=\sum _{j=1}^N a_j(x)e^{iS_j(x)/\hbar } \]
with the functions $S_j$ and $S_k$ weakly transverse
for any $j\ne k$.
If for any compact interval $K\subset J$,
we have
\[ \int _K |F|^2 (x) dx =o(1 ) \]
then all $a_j$'s  vanish identically.
\end{lemm}
If $P={\rm Op}(p)$ with $p \in C_o^\infty (T^\star J)$,
 using the $L^2$ $\hbar-$uniform continuity of $P$, 
we have 
\[ P F(x) =\sum _{j=1}^N  p(x, S'_j(x))a_j (x) e^{iS_j(x)/\hbar }=o(1)~. \]
One sees that the $a_j$'s vanish
by choosing $p$ in an appropriate way,
i.e. supported near a point $(x_0,S'_{j_0}(x_0))$.

\begin{lemm} \label{lemm:gutz}
Let us consider the distributions $D_k (\hbar )$ on $J_k$
defined by $D_k (\hbar )=\sum _{\gl _l (\hbar)\in J_k}
 \gd (\gl_l (\hbar))$,
then $D_k$ is microlocally in $T^\star J_k$
a locally finite sum of WKB functions $D_{j,l}$
associated to the Lagrangian manifolds
$t=lS'_j (y) $ with $j=1,\cdots, N_k $ and $l\in \Z $.
We have 
\[ D_{j,l}=\frac{1}{2\pi \hbar  }
e^{ilS_{j,\hbar}  (y)/\hbar}S'_{j,\hbar}(y) ~,\]
and
\[ D_{j,l}=\frac{(-1)^l}{2\pi \hbar  }e^{ilS_{j,O}  (y)/\hbar}
T_j(y)\left (1+ il\hbar S_{j,2}(y) + O(\hbar ^2)\right)~,\]
with $S_{j,\hbar} \equiv \sum _{k=0}^\infty \hbar^k S_{j,k}$ the
semi-classical
 actions associated to the $j-$th well.

\end{lemm}
This is a formulation of the semi-classical trace formula 
(see Appendix C).

\begin{lemm}
If $\mu_l (\hbar )$ is a semi-classical 
spectrum modulo $o(\hbar ^4)$
and $\tilde{D}_k (\hbar )=\sum _{\mu_l (\hbar)\in J_k}
 \gd (\mu_l (\hbar))$,
then, for any pseudo-differential operator
$P={\rm Op}_\hbar (p)$, with $p\in C_o^\infty (T^\star J_k )$,
we have
\[ \| P(D_k -\tilde{D}_k ) \|_{L^2(J_k)}=o(\hbar )~.\]
\end{lemm}
It is enough to prove it for
$p=\chi(E)\hat{\gr}(t)$ and then it is elementary
because $P\gd (\gl)=\hbar ^{-1}\chi (\gl)\gr ((y-\gl)/\hbar )$. 

From the three previous  Lemmas, it follows
that, with Assumption \ref{ss:sep},   the spectrum in $J_k$ modulo
$o(\hbar ^4)$ determine  the periods $T_{j}(y) $
and the actions $S_{j,2}(y)$.  
 
\subsection{Limit values of some integrals} \label{sec:limites}
Using the trick of Section \ref{sec:trick}, we can use
Abel's result (Section \ref{sec:abel}) once we know
the following limits (or asymptotic behaviours) as $y \ra E_j^+$
($j=0,1,\cdots $):
\begin{itemize}
\item $f^j_\pm (y) $
\item  $   \int _{H^{-1}(y)} {V''(x)} dt $  where  $H=\xi^2 + V(x)$ is the
classical Hamiltonian. Here  $H^{-1}(y)$ is oriented so that $dt >0$.
\item $f'^j _\pm (y)$
\end{itemize} 
All of them are determined by the knowledge of $V$ in the set
$\{ x | V(x) \leq E_j \} $. 

It is clear, except for the second one; we have:
\begin{lemm}\label{lemm:limites} Let us assume 
that $V$ satisfies Assumption 1 of Section \ref{sec:technical}.
If $E_j$ is a critical value of $V$ which is not
a local minimum  and 
 $\tau (z):=\int _{H^{-1}{(E_j+z)} } V''(x) dt - \int _{H^{-1}{(E_j-z)} } V''(x) dt$,
then $ \lim _{z\ra 0^+}\tau (z)=0$.
\end{lemm}
\begin{demo}
We cut the integrals into pieces. One piece near each critical
point and another piece far from them. 
Far from the critical points, the convergence is clear.
\begin{itemize}
\item {\it Local maximum:}
let us take a critical point where $V(x)=E_j -A(x-x_0)^{2N}(1+o(1))$
with $N\geq 1$ and $A>0$.
We use a smooth change of variable $x=\psi(y) $
with $\psi (0)=x_0$  so that
$V(\psi(y))=E_j -y^{2N}$.
We are reduced to check that
\[ \lim _{\ge \ra 0^+}\left( \int _{0}^1 \frac{W(y)dy}{\sqrt{\ge+y^{2N}}}
-  \int _{\ge^{1/2N}}^1 \frac{W(y)dy}{\sqrt{y^{2N}-\ge}} \right) =0 ~,\]
assuming that  $W(y)=O(y^{2N-2})$.
\item {\it Other critical points:} 
let us take a critical point where $V(x)=E_j +A(x-x_0)^{2N+1}(1+o(1))$
with $N\geq 1$ and $A>0$.
We use the same method.
\end{itemize}

\end{demo}
\section{Extensions to other operators}
\subsection{The statement}
Let us indicate in this Section how to extend the previous results
 to the
operator
 \[{ L}_{\hbar } = -\hbar ^2 \frac{d}{dx}
\left(n(x)\frac{d}{dx}\right) + n(x) ~\]
which was found in Section \ref{sec:acoustic}.
We want to recover the function $n(x)$.
Let us sketch the one well case for which we will get:
\begin{theo}
Assuming that
\begin{itemize} 
\item the function $n(x)$ admits a {\rm non degenerate minimum}
$n(x_0)=E_0>0 $, 
\item the function $n(x)$ has  {\rm no critical values 
    in $]E_0, E_1]$} with $E_1 \leq  \liminf _{x\ra \pa I}n(x)$,
\item
the function  $n(x)$ has {\rm a generic symmetry defect}
 as in Theorem
\ref{theo:main},
\end{itemize}
 then the function $n$ is determined in $\{ x|n(x)\leq E_1 \} $
by the semi-classical spectrum of $L_\hbar $ modulo
$o(\hbar^2)$.
\end{theo}
The proof works along the same lines as that of Theorem
\ref{theo:main} except that we get an integral  transform which is not exactly
Abel's transform. 

\subsection{The Weyl symbol and the actions}

The Weyl symbol $l$ of $L$ can be computed, using the Moyal product,  as
$l=\xi \star n \star\xi + n $.
We get:
\[ l(x,\xi)= n(x)(1+\xi ^2)+\frac{\hbar^2}{4}n''(x) ~. \]

The action $S_0$ satisfies:
\[ \frac{dS_0}{dy}(y)=T(y)=
\int _{n(x)\leq y}\frac{dx}{\sqrt{n(x)(y-n(x))}}~.\]

The action $S_2$ is given from \cite{CV2} by
\[ S_2(y)=-\frac{1}{12}\frac{d}{dy}\int_{\gg_y}\left( yn''-2
\left( \frac{y}{n}-1 \right)n'^2 \right)   dt -\frac{1}{4}\int_{\gg_y}
 n'' dt ~,\]
which we rewrite:
\[ S_2(y)=-\frac{1}{12}\frac{d}{dy}J(y)-\frac{1}{4}K(y)
  ~.\]
\begin{itemize}
\item {\bf The integral J:}
\[  J(y)= \int_{x_-(y)}^{x_+(y)}\left( yn''-
2\left( \frac{y}{n}-1 \right)n'^2 \right) \frac{dx}{\sqrt{n(y-n)}}\]

Using $x=f_\pm(y) $ as in Section \ref{sec:1well}
and 
\[ \Phi (y)=\frac{1}{f'_+(y)} -\frac{1}{f'_-(y)}~,\]
we get $J(y)=({\cal J}\Phi)(y)$, with 
\[  ( {\cal J} \Phi)(y)=
\int _{E_0}^y \left( y\Phi'(u) -2\left( \frac{y}{u}-1 \right)
\Phi(u) \right) \frac{du}{\sqrt{u(y-u)}}~.\]
\item {\bf The integral K:}
\[ K(y)=\int_{E_0}^y \Phi'(u)\frac{du}{\sqrt{u(y-u)}}\]
and
\[ K(y)=2\frac{d}{dy}\int_{E_0}^y \Phi'(u)\frac{\sqrt{y-u}\
  du}{\sqrt{u}}\]
which is rewritten as:
\[  K(y)=2\frac{d}{dy}({\cal K}\Phi)(y)\ .\]  
\end{itemize}

\subsection{An integral transform}
\begin{lemm} If $0<E_0<E_1$, 
the kernel of  $A:={\cal J}+6{\cal K}$
 on the space of continuous function on $[E_0,E_1]$
at most  two  dimensional and all functions in this kernel are smooth.
\end{lemm}
\begin{demo}
we have
\begin{equation} \label{equ:A}
 A\Phi (y)=\int _{E_0}^y\left(  (7y-6u)\Phi'(u) -2\left( \frac{y}{u}-1
\right) \Phi(u)\right) \frac{du}{\sqrt{u(y-u)}} \ .\end{equation}

We compute $T\circ A$ with  the operator $T$ defined
by
$T\psi (y)=\int _{E_0}^y \frac{\psi (u) du}{\sqrt{y-u}}$.
We will need the easy:
\begin{lemm}
We have:
\[ \int _{E_0}^y \frac{udu}{\sqrt{y-u}}\int _{E_0}^u  f (t) 
\frac{dt}{\sqrt{u-t}}= \frac{\pi}{2} 
\int  _{E_0}^y (t + y)f (t) dt ~,\]
and 
\[ \int _{E_0}^y \frac{du}{\sqrt{y-u}}\int _{E_0}^u  f (t) 
\frac{dt}{\sqrt{u-t}}= \pi 
\int  _{E_0}^y f (t) dt ~,\]
\end{lemm}

Applying the previous formulae, we get:
\[ T\circ A\  (\Phi) (y)=
\frac{\pi}{2} \int_{E_0}^y \lbrack (t+y)(7\Phi'(t)-2\frac{\Phi(t)}{t})
+2 (-6t\Phi'(t)+2   \Phi(t) ) \rbrack \frac{dt}{\sqrt{t}}~. \]
Taking two derivatives:
\[ \frac{\pi}{y^{3/2}}\frac{d^2}{dy^2}((T\circ A) \ \Phi) (y)=
 y^2 \Phi'' (y) +4y \Phi'(y) - \Phi(y)
 ~.\]

From $S_2$ and $A \Phi (E_0)$, we get $A\Phi $, then 
we get $P(\Phi )$ where $P\phi =y^2\phi''+4y\phi' -\phi $
 is a non singular linear differential
equation (remind that $E_0>0$). So,  if we know also
 $\Phi (E_0)$ and the asymptotic behaviour
of $\Phi'(E_0)$, we can get $\Phi $.
Let us assume $n''(x_0)=a>0$.
Then we have:
\begin{itemize}
\item $A\Phi (E_0)=2\pi \sqrt{aE_0}$
\item $\Phi(E_0)=0$
\item $\Phi'(y)\sim 4\sqrt{a}/\sqrt{y-E_0} $.
\end{itemize}
\end{demo}
%

\section*{Appendix A: Abel's result} \label{sec:abel}
 
Let us consider the linear operator
$T$ which acts on continuous functions on $[E_0,E [$
defined by:
\[ Tf(x)=\int _{E_0}^x \frac{f(y)dy}{\sqrt{x-y}}~.\]
Then $T^2 f(x)  =\pi \int_{E_0}^x f(y)dy $.
This implies that $T$ is injective!
This is the content of \cite{abel}.

\section*{Appendix B: a proof of the $\Psi DO$ trace
 formula of Section \ref{sec:bs}}
For this Section, one can read \cite{Gracia}.
This can be seen as a complement
and a partial rewriting  of my paper \cite{CV2} with a better trace
formula.
The formula we will prove is more general than that in Section
\ref{sec:bs}. It is valid even for several wells. Let us state it:
\begin{theo}
Let $f\in C_o^\infty (J_k)$
 and $F(y):=-\int _y^\infty f(u)du $,
we have, with $Z=T^\star I$, modulo $ O(\hbar ^\infty)$: 
\[ {\rm Trace}F(\hat{H})\equiv \frac{1}{2\pi \hbar }
\left( \int _{Z}F(H) dx d\xi 
+  {\hbar ^2}\int _{J_k} f(y)\left(\sum _{j=1}^{N_k} (S_{2,j}^k (y)+
\hbar^2 S_{4,j}^k(y)+
 \cdots)\right) dy  \right) ~.\]
\end{theo}
\begin{demo}
\begin{enumerate}
\item {\it Reduction to $N_k=1$:}
we can decompose both the lefthandside and the righthandside
according to the $N_k$ wells: for the lhs, it uses the fact that the 
classical spectrum splits into $N_k$ parts; for the rhs, it is enough
to decompose the first integral terms according to the connected 
component of $H< E_k$.
\item {\it Reduction from $N_k=1$ to one well:}
the whole Moyal symbol of $F(\hat{H})$ is $\equiv F(E_0)$ 
in $\{ H\leq E_{k-1} \} $.
\item {\it The harmonic oscillator case ($\hat{H}=\Omega $): }
\[ {\rm Trace}F(\Omega )=\ha \sum _{n\in \Z} \tilde{F}\left( (n+\ha)\hbar
\right) ~\]
with $\tilde{F}$ even and co\"{i}nciding with $F$ on the positive axis.
We get with Poisson summation formula:
\[ {\rm Trace}F(\Omega )=\frac{1}{2\pi \hbar } \int\!\int F\left( 
\frac{x^2 +\xi ^2}{2} \right) dx d\xi  +O(\hbar ^\infty)~.\]
\item {\it The case where $F$ is compactly supported:}
using Poisson summation formula as in \cite{CV2}, we get
\[ {\rm Trace}F (\hat{H})=\frac{1}{2\pi \hbar } 
\int F(y)S'(y) dy \]
and we get this case by integration by part.
\item {\it The final step:}
we can assume that $H=\frac{(x-x_0)^2 +\xi ^2}{2} +E_0$ 
near $(x_0,0)$ and we split $F=F_0+F_1$ where 
\[ F_0(H)\equiv F_1 \left(\frac{(x-x_0)^2 +\xi ^2}{2} +E_0\right)~.\]
The formula then follows from the two  particular cases computed before.
\end{enumerate}
\end{demo}

{\it For the convenience of the reader, we regive also the way to get 
$S_2$ from the Moyal formula.}

Defining
$F^\star (H)$ by
$F(\hat{H})={\rm Op}_{\rm Weyl} (F^\star (H))$
we know that, with $z_0=(x_0,\xi_0)$ and $H_0=H(z_0)$,
\[ F^\star (H)(z_0)=F(H_0)+\ha F''(H_0)(H-H_0)^{\star 2 }(z_0)+
\frac{1}{6} F'''(H_0)(H-H_0)^{\star 3 }(z_0) + O(\hbar^4) ~.\]
Computing the Moyal powers of $H-H_0$ at the point $z_0$
mod $  O(\hbar^4)$, gives
\[ F^\star (H)=F(H)-\hbar^2 \left(\frac{1}{8}f'(H){\rm det}(H'')
   +\frac{1}{24} f''(H)H''(X_H,X_H)     \right)+  O(\hbar^4)~.\]
If  $\ga=
\iota (X_H) H'' $, we have
$d\ga= 2{\rm det} (H'') d\xi \wedge dx $,
and 
we get,
by Stokes and with $\gg_y$ oriented according to the dynamics:
\[ \int _{\gg_y} \ga = 2\int _{H\leq y}{\rm
  det}(H'')dxd\xi \]
and the final result for $S_2(y)$ using an integration by part
and the formula $dtdy=dxd\xi $:
  \[ S_2(y)=
-1/24 \int_{\gg_y}{\rm det}(H'') dt ~.\]

\section*{Appendix C: the 
semi-classical trace formula} \label{sec:gutz}
In this Section, we want to give a proof of Lemma \ref{lemm:gutz}.

We want to evaluate mod $O(\hbar ^\infty )$
the sums:
\[ D(y):=\frac{1}{\hbar}\sum _{l\in \Z}
\gr \left( \frac{y-S^{-1}(2\pi l  \hbar)}{\hbar } \right) ~,\]
where $S: \R \ra \R $ is an extension to $\R$  of the given function 
$S_j$ on $\Delta $ which is $\equiv {\rm Id}$ near infinity.
This is the equal to $D_{\Delta, \gr}^j (y)$ up to   $O(\hbar ^\infty
)$.
Using the Poisson summation formula and defining
\[ F_y(x)=
\int _\R \gr  \left( \frac{y-S^{-1}(\hbar y))}{\hbar } \right)e^{-ixy}
dy ~,\]
we get 
\begin{equation} \label{equ:PSF}
 D(y)= \frac{1}{2 \pi \hbar} \sum _{m\in \Z} F_y(m) ~.\end{equation}
Using the change of variable,
$y-S^{-1}(\hbar y)=\hbar z$ or $y=S(y-\hbar z )/\hbar $, we get:
\[ F_y(x)=\int \gr (z)e^{-ixS(y-\hbar z)/\hbar }
S'(y-\hbar z) dz ~.\]
Using the fact that all moments of  $\gr$ vanish
and Taylor expanding $S(y-\hbar z)$ w.r. to $\hbar$, we get
\[ F_y(x)=e^{-ixS(y)/\hbar }S'(y) \hat{\gr}(-xS'(y))
+O(\hbar^\infty)~.\]
If the support of $\hat{\gr}$ is close enough to $S'(y)$, we get
the final answer taking the contribution of $m=-1$ to 
Equation (\ref{equ:PSF}). This way, we get the formula
of Lemma \ref{lemm:gutz}.

\bibliographystyle{plain}


\end{document}

%% file: potential.pstex_t
\begin{picture}(0,0)%
\includegraphics{potential.pstex}%
\end{picture}%
\setlength{\unitlength}{4144sp}%
\begingroup\makeatletter\ifx\SetFigFont\undefined%
\gdef\SetFigFont#1#2#3#4#5{%
  \reset@font\fontsize{#1}{#2pt}%
  \fontfamily{#3}\fontseries{#4}\fontshape{#5}%
  \selectfont}%
\fi\endgroup%
\begin{picture}(4614,2246)(169,-1497)
\put(1936,187){\makebox(0,0)[lb]{\smash{{\SetFigFont{12}{14.4}{\rmdefault}{\mddefault}{\updefault}{\color[rgb]{0,0,0}$y=V(x)$}%
}}}}
\put(901,299){\makebox(0,0)[lb]{\smash{{\SetFigFont{12}{14.4}{\rmdefault}{\mddefault}{\updefault}{\color[rgb]{0,0,0}$y$}%
}}}}
\put(2161,-1321){\makebox(0,0)[lb]{\smash{{\SetFigFont{12}{14.4}{\rmdefault}{\mddefault}{\updefault}{\color[rgb]{0,0,0}$x$}%
}}}}
\put(2843,473){\makebox(0,0)[lb]{\smash{{\SetFigFont{12}{14.4}{\rmdefault}{\mddefault}{\updefault}{\color[rgb]{0,0,0}$x$}%
}}}}
\put(4643,-990){\makebox(0,0)[lb]{\smash{{\SetFigFont{12}{14.4}{\rmdefault}{\mddefault}{\updefault}{\color[rgb]{0,0,0}$y$}%
}}}}
\put(4027,254){\makebox(0,0)[lb]{\smash{{\SetFigFont{12}{14.4}{\rmdefault}{\mddefault}{\updefault}{\color[rgb]{0,0,0}$f_+(y)$}%
}}}}
\put(2852,-743){\makebox(0,0)[lb]{\smash{{\SetFigFont{12}{14.4}{\rmdefault}{\mddefault}{\updefault}{\color[rgb]{0,0,0}$x_0$}%
}}}}
\put(4028,-1291){\makebox(0,0)[lb]{\smash{{\SetFigFont{12}{14.4}{\rmdefault}{\mddefault}{\updefault}{\color[rgb]{0,0,0}$f_-(y)$}%
}}}}
\put(977,-1314){\makebox(0,0)[lb]{\smash{{\SetFigFont{12}{14.4}{\rmdefault}{\mddefault}{\updefault}{\color[rgb]{0,0,0}$x_0$}%
}}}}
\put(3144,-1132){\makebox(0,0)[lb]{\smash{{\SetFigFont{12}{14.4}{\rmdefault}{\mddefault}{\updefault}{\color[rgb]{0,0,0}$E_0$}%
}}}}
\put(676,-908){\makebox(0,0)[lb]{\smash{{\SetFigFont{12}{14.4}{\rmdefault}{\mddefault}{\updefault}{\color[rgb]{0,0,0}$E_0$}%
}}}}
\end{picture}%

%% file: contrex.pstex_t
\begin{picture}(0,0)%
\includegraphics{contrex.pstex}%
\end{picture}%
\setlength{\unitlength}{4144sp}%
\begingroup\makeatletter\ifx\SetFigFont\undefined%
\gdef\SetFigFont#1#2#3#4#5{%
  \reset@font\fontsize{#1}{#2pt}%
  \fontfamily{#3}\fontseries{#4}\fontshape{#5}%
  \selectfont}%
\fi\endgroup%
\begin{picture}(4734,3329)(424,-2608)
\put(2318,-1411){\makebox(0,0)[lb]{\smash{{\SetFigFont{12}{14.4}{\rmdefault}{\mddefault}{\updefault}{\color[rgb]{0,0,0}$I$}%
}}}}
\put(2296,-833){\makebox(0,0)[lb]{\smash{{\SetFigFont{12}{14.4}{\rmdefault}{\mddefault}{\updefault}{\color[rgb]{0,0,0}$II$}%
}}}}
\put(2304,-53){\makebox(0,0)[lb]{\smash{{\SetFigFont{12}{14.4}{\rmdefault}{\mddefault}{\updefault}{\color[rgb]{0,0,0}$III$}%
}}}}
\put(2101,562){\makebox(0,0)[lb]{\smash{{\SetFigFont{12}{14.4}{\rmdefault}{\mddefault}{\updefault}{\color[rgb]{0,0,0}$V(x)$}%
}}}}
\put(4066,-1778){\makebox(0,0)[lb]{\smash{{\SetFigFont{12}{14.4}{\rmdefault}{\mddefault}{\updefault}{\color[rgb]{0,0,0}$x$}%
}}}}
\end{picture}%

%% file: proof.pstex_t
\begin{picture}(0,0)%
\includegraphics{proof.pstex}%
\end{picture}%
\setlength{\unitlength}{4144sp}%
\begingroup\makeatletter\ifx\SetFigFont\undefined%
\gdef\SetFigFont#1#2#3#4#5{%
  \reset@font\fontsize{#1}{#2pt}%
  \fontfamily{#3}\fontseries{#4}\fontshape{#5}%
  \selectfont}%
\fi\endgroup%
\begin{picture}(7476,1466)(219,-1425)
\put(308,-593){\makebox(0,0)[lb]{\smash{{\SetFigFont{12}{14.4}{\rmdefault}{\mddefault}{\updefault}{\color[rgb]{0,0,0}Semi-classical }%
}}}}
\put(308,-818){\makebox(0,0)[lb]{\smash{{\SetFigFont{12}{14.4}{\rmdefault}{\mddefault}{\updefault}{\color[rgb]{0,0,0}spectrum mod}%
}}}}
\put(308,-1043){\makebox(0,0)[lb]{\smash{{\SetFigFont{12}{14.4}{\rmdefault}{\mddefault}{\updefault}{\color[rgb]{0,0,0}$o(\hbar^2)$}%
}}}}
\put(2386,-1186){\makebox(0,0)[lb]{\smash{{\SetFigFont{12}{14.4}{\rmdefault}{\mddefault}{\updefault}{\color[rgb]{0,0,0}$E_0,~V''(x_0)$}%
}}}}
\put(4343,-653){\makebox(0,0)[lb]{\smash{{\SetFigFont{12}{14.4}{\rmdefault}{\mddefault}{\updefault}{\color[rgb]{0,0,0}$I,~J$}%
}}}}
\put(5731,-534){\makebox(0,0)[lb]{\smash{{\SetFigFont{12}{14.4}{\rmdefault}{\mddefault}{\updefault}{\color[rgb]{0,0,0}$\pm F', G'$}%
}}}}
\put(5041,-1336){\makebox(0,0)[lb]{\smash{{\SetFigFont{12}{14.4}{\rmdefault}{\mddefault}{\updefault}{\color[rgb]{0,0,0}V up to a symmetry-translation}%
}}}}
\put(5011,-443){\makebox(0,0)[lb]{\smash{{\SetFigFont{12}{14.4}{\rmdefault}{\mddefault}{\updefault}{\color[rgb]{0,0,0}Abel}%
}}}}
\put(1868,-901){\makebox(0,0)[lb]{\smash{{\SetFigFont{12}{14.4}{\rmdefault}{\mddefault}{\updefault}{\color[rgb]{0,0,0}$\gl_1$}%
}}}}
\put(1703,-504){\makebox(0,0)[lb]{\smash{{\SetFigFont{12}{14.4}{\rmdefault}{\mddefault}{\updefault}{\color[rgb]{0,0,0}$\Psi DO$ trace formula}%
}}}}
\put(2513,-136){\makebox(0,0)[lb]{\smash{{\SetFigFont{12}{14.4}{\rmdefault}{\mddefault}{\updefault}{\color[rgb]{0,0,0}$T(y),~S_2(y)$}%
}}}}
\end{picture}%

%% file: pot3.pstex_t
\begin{picture}(0,0)%
\includegraphics{pot3.pstex}%
\end{picture}%
\setlength{\unitlength}{4144sp}%
\begingroup\makeatletter\ifx\SetFigFont\undefined%
\gdef\SetFigFont#1#2#3#4#5{%
  \reset@font\fontsize{#1}{#2pt}%
  \fontfamily{#3}\fontseries{#4}\fontshape{#5}%
  \selectfont}%
\fi\endgroup%
\begin{picture}(4648,2761)(270,-1990)
\put(3316,-1748){\makebox(0,0)[lb]{\smash{{\SetFigFont{12}{14.4}{\rmdefault}{\mddefault}{\updefault}{\color[rgb]{0,0,0}$I$}%
}}}}
\put(3278,-1418){\makebox(0,0)[lb]{\smash{{\SetFigFont{12}{14.4}{\rmdefault}{\mddefault}{\updefault}{\color[rgb]{0,0,0}$II$}%
}}}}
\put(3233,-848){\makebox(0,0)[lb]{\smash{{\SetFigFont{12}{14.4}{\rmdefault}{\mddefault}{\updefault}{\color[rgb]{0,0,0}$III$}%
}}}}
\put(4861,-1808){\makebox(0,0)[lb]{\smash{{\SetFigFont{12}{14.4}{\rmdefault}{\mddefault}{\updefault}{\color[rgb]{0,0,0}$x$}%
}}}}
\put(316,-1621){\makebox(0,0)[lb]{\smash{{\SetFigFont{12}{14.4}{\rmdefault}{\mddefault}{\updefault}{\color[rgb]{0,0,0}$E_1$}%
}}}}
\put(302,-1245){\makebox(0,0)[lb]{\smash{{\SetFigFont{12}{14.4}{\rmdefault}{\mddefault}{\updefault}{\color[rgb]{0,0,0}$E_2$}%
}}}}
\put(285,-195){\makebox(0,0)[lb]{\smash{{\SetFigFont{12}{14.4}{\rmdefault}{\mddefault}{\updefault}{\color[rgb]{0,0,0}$E_\infty $}%
}}}}
\put(647,577){\makebox(0,0)[lb]{\smash{{\SetFigFont{12}{14.4}{\rmdefault}{\mddefault}{\updefault}{\color[rgb]{0,0,0}$V(x)$}%
}}}}
\put(1426,-1396){\makebox(0,0)[lb]{\smash{{\SetFigFont{12}{14.4}{\rmdefault}{\mddefault}{\updefault}{\color[rgb]{0,0,0}$-$}%
}}}}
\put(2566,-1381){\makebox(0,0)[lb]{\smash{{\SetFigFont{12}{14.4}{\rmdefault}{\mddefault}{\updefault}{\color[rgb]{0,0,0}$+$}%
}}}}
\put(293,-1921){\makebox(0,0)[lb]{\smash{{\SetFigFont{12}{14.4}{\rmdefault}{\mddefault}{\updefault}{\color[rgb]{0,0,0}$E_0$}%
}}}}
\end{picture}%

%% file: period.pstex_t
\begin{picture}(0,0)%
\includegraphics{period.pstex}%
\end{picture}%
\setlength{\unitlength}{4144sp}%
\begingroup\makeatletter\ifx\SetFigFont\undefined%
\gdef\SetFigFont#1#2#3#4#5{%
  \reset@font\fontsize{#1}{#2pt}%
  \fontfamily{#3}\fontseries{#4}\fontshape{#5}%
  \selectfont}%
\fi\endgroup%
\begin{picture}(3709,2349)(391,-1888)
\put(593,-1823){\makebox(0,0)[lb]{\smash{{\SetFigFont{12}{14.4}{\rmdefault}{\mddefault}{\updefault}{\color[rgb]{0,0,0}$0$}%
}}}}
\put(2993,-1815){\makebox(0,0)[lb]{\smash{{\SetFigFont{12}{14.4}{\rmdefault}{\mddefault}{\updefault}{\color[rgb]{0,0,0}$E_\infty$ }%
}}}}
\put(406,314){\makebox(0,0)[lb]{\smash{{\SetFigFont{12}{14.4}{\rmdefault}{\mddefault}{\updefault}{\color[rgb]{0,0,0}$T$}%
}}}}
\put(4028,-1613){\makebox(0,0)[lb]{\smash{{\SetFigFont{12}{14.4}{\rmdefault}{\mddefault}{\updefault}{\color[rgb]{0,0,0}$E$}%
}}}}
\put(1598,-1817){\makebox(0,0)[lb]{\smash{{\SetFigFont{12}{14.4}{\rmdefault}{\mddefault}{\updefault}{\color[rgb]{0,0,0}$E_1$}%
}}}}
\put(2250,-1816){\makebox(0,0)[lb]{\smash{{\SetFigFont{12}{14.4}{\rmdefault}{\mddefault}{\updefault}{\color[rgb]{0,0,0}$E_2$}%
}}}}
\put(1764,-1395){\makebox(0,0)[lb]{\smash{{\SetFigFont{12}{14.4}{\rmdefault}{\mddefault}{\updefault}{\color[rgb]{0,0,0}$T_-(E)$}%
}}}}
\put(781,-1291){\makebox(0,0)[lb]{\smash{{\SetFigFont{12}{14.4}{\rmdefault}{\mddefault}{\updefault}{\color[rgb]{0,0,0}$T_+(E)$}%
}}}}
\end{picture}%